\documentclass[showpacs,showkeys,preprintnumbers,amsmath,amssymb,times,graphicx]{revtex4}

\usepackage{graphicx} 
\usepackage{dcolumn}
\usepackage{bm}

\usepackage{color}
\usepackage{amsmath}
\usepackage{amsfonts}
\usepackage{amssymb}
\usepackage{amsbsy}
\usepackage[figuresright]{rotating}
\usepackage[font=small,labelfont=bf]{caption}
\usepackage{natbib}

\def\3{\ss}

\def\q0{\phantom{1}}

\def\thetabn{\Theta_{Bn}}

\def\parp2{\frac{\partial^{2}}{\partial p^{2} }}

\def\m21{$2^{\circ}\times 1^{\circ}$}

\def\ts{\thinspace}

\def\ne8{Ne\ts{$\scriptstyle {\rm VIII}$} }

\newcommand{\asr}{{Adv. \ Space. \ Res.}}

\newcommand{\grl}{{Geo\-phys.\ Res.\ Lett.}}

\newcommand{\jgr}{{J.\ Geo\-phys.\ Res.}}

\newcommand{\ssr}{{Space Sci.\,Rev.}}

\newcommand{\pf}{{Phys. Fluids}}

\newcommand{\pop}{{Phys.\ Plasmas}}

\def\ion[#1 #2]{#1\,{\sc #2}}
\def\lamb[#1]{#1\,{\AA}}
\def\serts89{SERTS-89}
\def\fe12{Fe\,{\sc xii}}

\def\mb[#1]{\makebox[0.15cm][l]{#1}}

\begin{document}




\title{Fundamentals of  Non-relativistic Collisionless Shock Physics: \\  I. The Shock Problem}

\author{R. A. Treumann$^\dag$ and C. H. Jaroschek$^{*}$}\email{treumann@issibern.ch}
\affiliation{$^\dag$ Department of Geophysics and Environmental Sciences, Munich University, D-80333 Munich, Germany  \\ 
Department of Physics and Astronomy, Dartmouth College, Hanover, 03755 NH, USA \\ 
$^{*}$Department Earth \& Planetary Science, University of Tokyo, Tokyo, Japan
}%

\begin{abstract}
The problem of collisionless shocks is posed as the problem of understanding how in a completely collisionless streaming high-temperature plasma shocks can develop at all, forming discontinuous transition layers of thickness much less than any collisional mean free path length. The history of shock research is briefly reviewed. It is expressed that collisionless shocks as a realistic possibility of a state of matter have been realized not earlier than roughly half a centruy ago. The basic properties of collisionless shocks are noted in preparing the theory of collisionless shocks and a classification of shocks is given in terms of their physical properties, which is developed in the following chapters. The structure of this chapter is as follows: 1. A cursory historical overview, describing the early history, gasdynamic shocks, the realisation of the existence of collisionless shocks and their investigation over three decades in theory and observation until the numerical simulation age, 2. Posing the shock problem as the question: When are shocks? 3. Types of collisionless shocks, describing electrostatic shocks, magnetized shocks, MHD shocks, shock evolutionarity and coplanarity, switch-on and switch-off shocks, 4. Criticality, describing the transition from subcritical dissipative to supercritical viscose shocks, 5. Remarks.
\end{abstract}
\maketitle 
\section{Introduction}
\noindent Collisionless shocks are shock waves that evolve in dilute hot gases, hot enough for binary collisions to become completely unimportant over the shock width. Such gases are therefore in the state of fully ionised plasmas and, under normal conditions, do not exist on Earth. However, they are abundant in space. 

As long as collisionless shocks are nonrelativistic, they are invisible from remote. They may emit low-frequency radio waves but do not shine, neither in the visible light nor in higher energy radiation like x-rays or gamma-rays. The shocks we see in astrophysical objects are relativistic shocks which follow quite a different physics than non-relativistic shocks. Examples of such shocks are well known from supernova remnants where they produce the spider web forms in the expanding gas. They are also known from astrophysical jets which are generated from the central engine in supernovae, active galactic nuclei and black holes. These relativistic shocks are visible: they emit all kinds of radiation, radio waves, infrared light, visible light, x-rays and, in some cases, even gamma-rays. They are probably  also responsible for the generation of the most energetic cosmic rays.

Non-relativistic shocks, on the other hand, even though they are invisible `dark' objects, constitute most probably the absolute majority among shocks in cosmic space in the universe. Their responsibilities are spanned wide, from plasma heating, through the generation of dissipation, up to the acceleration of particles into medium energies, which may also serve as the seed particle distribution that can be further accelerated by relativistic shocks to become energetic cosmic rays. Non-relativistic shocks have, therefore, been of substantial interest for more than half a century. Collisional shocks have in the past been treated in depth in prominent reviews \citep[cf., e.g.,][]{Zeldovich1966} in the gasdynamic approach. However, it has turned out that most of the shocks in space and in the cosmos are in fact completely non-collisional with processes at the smallest plasma scales dominating their physics. Their proper description requires the application of kinetic theory. Understanding the physics of non-relativistic collisionless shocks is also a pre-requisite for the development of a proper theory of relativistic shocks. In view of this perspective the present series of reviews tries to provide an overview over what has been achieved until today in the physics of non-relativistic shocks.

\section{A Cursory Historical Overview}\label{intro-1}
\noindent The present series of papers deals with the physics of shock waves in our heliosphere only, a very particular class of shocks: shocks in collisionless high temperature but non-relativistic plasmas. The physics of shocks waves is however much more general, covering one of the most interesting chapters in physics, from solid state to the huge dimensions of cosmic space. Shock waves were involved when stars and planets formed and when the matter in the universe clumped to build galaxies, and they are involved when the heavy elements have formed which are at the fundament of life and civilisation. Mankind, however, has become aware of shocks only very lately. The present section gives a short account of its history in human understanding.
\subsection{The early history}
\begin{figure}[t!]
\hspace{0.0cm}\centerline{\includegraphics[width=0.75\textwidth,clip=]{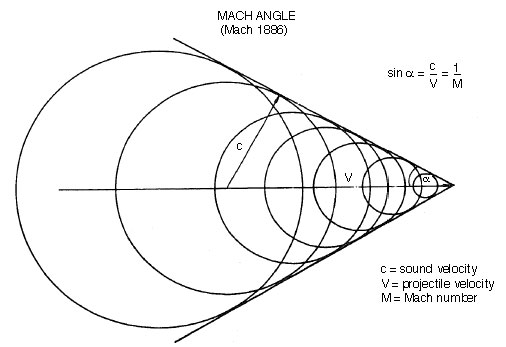} }
\caption[1]
{\footnotesize Mach's 1886 drawing of the form of a projectile generated shock of Mach-opening angle $\alpha=\sin^{-1}(1/{\cal M})$. The projectile is moving to the right (or the air is moving to the left) at speed $V$. The shock front is formed as the envelope of all the spherical soundwave fronts excited along the path of the projectile in the air when the sound waves expand radially at sound speed $c$.}\label{Mach-f1}
\end{figure}

\noindent Interest in shocks has arisen first in gasdynamics when fast flows came to attention not long before \index{Mach, Ernst} Ernst Mach's realization \citep{Mach1884,Mach1886,Mach1887,Mach1898} that in order for a shock in a flow to evolve an obstacle must be put into the flow with the property that the relative velocity of the obstacle $V$ with respect to the bulk flow exceeds the velocity of sound $c_s$ in the medium, leading him to define the critical velocity ratio ${\cal M}=V/c_s$ (see Figure \ref{Mach-f1}), known today as the famous \index{Mach number} Mach number and contributing even more to his public recognition and scientific immortality than his other great contribution to physics, the preparation of the path to General Relativity for Albert Einstein by his fundamental analysis of the nature of gravity and equivalence.

\begin{figure}[t!]
\hspace{0.0cm}\centerline{\includegraphics[width=0.5\textwidth,clip=]{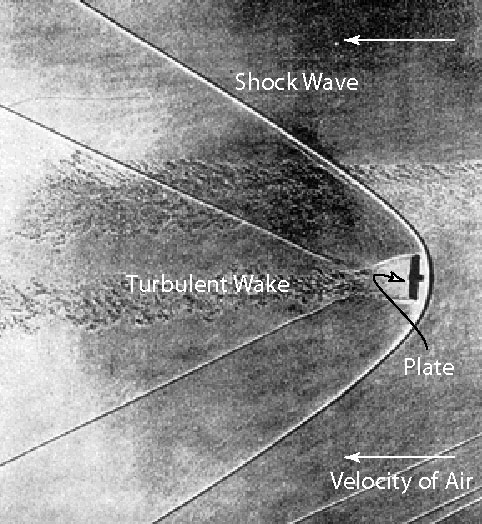} }
\caption[1]
{\footnotesize A cord photograph of a blunt plate moving at supersonic velocity in a transparent viscous medium causing a blunt thin shock wave of hyperbolic form and leaving a turbulent wake behind (photograph taken after H. Schardin, Lilienthal-Gesellschaft, Report No. 139). This figure shows nicely that blunt objects cause blunt nosed shocks standing at a distance from the object. This is similar to what the Earth's dipole magnetic field does in the solar wind.}\label{Mach-f2}
\end{figure}

Mach was experimenting at this time with projectiles that were ejected from guns using the new technique of taking photographs of cords generated by projectiles when traversing a transparent though very viscous fluid (for an example see Figure \ref{Mach-f2}). Even though Mach was not working for the Austrian Ministry of War, these endeavours were in the very interest of it.  Mach mentioned this fact, somewhat ironically, in his famous later publication \citep{Mach1898}  where he was publicly reviewing his results on this matter and where he writes: ``As in our todays life shooting and everything connected to it under certain circumstances plays a very important, if not the dominant, role you may possibly turn your interest for an hour to certain experiments, which have been performed not necessarily in view of their martial application but rather in scientific purpose, and which will provide you some insight into the processes taking place in the shooting"  \citep{Pohl2002}.

The physical interpretation of Mach's observations was stimulated by the recognition that no matter how fast the stream would be, as seen from a reference frame at rest with the projectile, the projectile being an obstacle in the flow causes disturbances to evolve in the flow. Such disturbances are travelling waves which in ordinary gasdynamics are sound waves. In flows with Mach number ${\cal M}<1$ less than one sound waves can reach any upstream position thereby informing the flow of the presence of the obstacle and leaving the flow sufficient time for rearranging and changing its direction in order of turning smoothly around the obstacle. In this case no shock will evolve. 

In the opposite case ${\cal M}\geq 1$ the stream remains uninformed until it hits the obstacle completely unprepared for its presence, and something catastrophic happens, i.e. {\it the flow is shocked}. In this case the flow is too fast for the disturbances generated by the obstacle to propagate large distances upstream. They can propagate only a finite distance from the obstacle up to a certain point in a time that is shorter than the flow needs from this point to arrive at the obstacle. Hence it is clear that from the farthest upstream position reached by the waves --  i.e. from the shock to the obstacle -- the flow velocity must have dropped below the velocity of sound, which requires the shock to be a thin discontinuity surface at which the flow has been braked and a substantial part of its directed motional energy is transformed into heat, i.e. into disordered motion. The shock discontinuity surface thus separates the undisturbed cooler upstream flow from the disturbed warmer downstream flow between shock and obstacle that contains all the obstacle-excited sound waves, and the temperature and density of the flow must both increase across the shock from upstream to downstream which implies that the pressure, which is the product of density and temperature, increases as well. 

\subsection{\index{shocks}\index{shocks!gasdynamic}Gasdynamic shocks}
\noindent The geometry of the shock \index{shocks!geometry}can be constructed from the characteristic curves of sound wave propagation around the obstacle under the assumption that the shock is considered to be thin with respect to the wavelength of the sound. In the downstream region the subsonic flow bends around the obstacle. Since  retardation implies that the volume of a flow element is reduced, the downstream density is increased above upstream and flow energy is converted into thermal energy giving the region downstream of the shock  a higher temperature than upstream. Higher temperature means increased disorder and thus  implies that entropy is generated at the shock making the whole process irreversible, which can be done only when the shock front supports dissipation. Finally, due to the presence of the broad spectrum of sound waves the downstream region supports irregular motion and is to some extent turbulent. 

This is the simple global physics implied. It can be described by the equations of compressible gasdynamics where the emphasis is on compressible since sound waves are fluctuations in density. The microscopic processes are, however, more complicated even in ordinary gasdynamics implying some knowledge about heat conduction and viscosity which determine the physical thickness of the shock front, the shock profile, and the process of heating and generation of entropy. A first one dimensional theory goes back to  \cite{Becker1922}. Later comprehensive reviews can be found in \cite{Landau1959} and  \cite{Zeldovich1966}.  An extended review of the history of shock waves has been given only recently by  \cite{Krehl2007}. \index{Becker, Richard}\index{Landau, Leo D.}\index{Zeldovich, Yaakov B.}

Gasdynamic shocks are dominated by binary collisions and thus are collisional.  Binary collisions between the molecules of the fluid respectively the gas are required since collisions are the only way the molecules interact among each other as long as radiative interactions are absent. They are required for the necessary heating and entropy generation. There is also some dissipative interaction with the sound waves; this is, however, weak and usually negligible compared to the viscous interaction except under conditions when the amplitudes of the sound waves at the shortest wave lengths become large. This happens to be the case only inside the shock front the width of which is of the order of just a few \index{mean free path!collisional}collisional mean free paths $\lambda_{\rm mfp}=(n\sigma_c)^{-1}$. The latter is defined as the inverse product of gas number density $n$ and \index{cross section!collisional}collisional cross section $\sigma_c\simeq {\rm a\, few }\,\times 10^{-19}$\,m$^{-2}$. Thus for the mean free path being small requires very large gas densities which are rather rarely found in interstellar or interplanetary space, the subject we are interested in this book. Shocks developing in such practically collisionless conditions are called collisionless shocks, implying that binary collisions can completely be neglected. Their widths are much less than the theoretical collisional mean free path, and any dissipative processes must be attributed to mechanisms based not on collisions but on collective processes.

In gases collisionless conditions evolve naturally when the temperature of the gas rises. In this case the gas becomes dilute with decreasing density, and in addition the (generally weak) dependence of the collisional cross section on temperature causes a decrease in $\sigma_c$. This effect becomes particularly remarkable when the temperature substantially exceeds the ionisation energy threshold. Then the gas makes the transition to a plasma consisting of an ever decreasing number of neutral molecules and an increasing number of electron-ion pairs. At temperatures far above \index{energy!of ionization}ionisation energy the density of neutrals can be neglected compared to the density of electrons and ions. These are electrically charged particles with the interaction between them dominated not anymore by short range collisions but by the long range Coulomb force that decays $\propto r^{-2}$ with interparticle distance $r=|{\bf r}_2-{\bf r}_1|$. The collisional mean free path $\lambda_{\rm mfp}=(n\sigma_{\rm C})^{-1}\equiv \lambda_{\rm C}$ now contains the \index{cross section!Coulomb}Coulomb-cross section $\sigma_c=\sigma_{\rm C}$ thereby becoming \index{mean free path!Coulomb}the Coulomb mean free path  $\lambda_{\rm C}$. In plasma the two components electrons and ions are about independent populations coupled primarily through the condition of maintaining the quasi-neutrality of the plasma. The Coulomb cross section itself is inversely proportional to the fourth power of the particle velocity, and $\lambda_{\rm C}\sim v^4$ for fast particles increases rapidly, readily becoming as large as the macroscopic extension of the entire plasma. For thermal particles it increases as the square of the temperature, $\lambda_{\rm C}\propto T^2$, thus becoming as well very large. In interplanetary space this length is of the order of several AU. Any shock of lesser width must therefore be completely collisionless. This inescapable conclusion poses some severe problems in the interpretation and physical understanding of collisionless shocks. It is such shocks which are frequently encountered in interplanetary and interstellar space and which behave completely different from their collisional counterparts in fluids and gases. The present book deals with their properties as inferred from direct observation and interpreted in theory and simulation.

\subsection{Realizing collisionless shocks}
\noindent The possibility of collisionless shocks in an ionized gaseous medium that can be described by the equations of magnetohydrodynamics has been first anticipated by \cite{Courant1948}.  The first theory of magnetized shocks was given by \cite{DeHoffman1950} in an important paper that was stimulated by after-war atmospheric nuclear explosions. This paper was even preceded by Fermi's seminal paper on the origin of \index{acceleration!cosmic ray}cosmic rays \citep{Fermi1949} where he implicitly proposed the existence of collisionless shocks when stating that cosmic rays are accelerated in \index{Fermi, Enrico}multiple head-on collisions with magnetized shock waves  in each collision picking up the difference in flow velocity between the flows upstream and downstream of the shock. The De Hoffman \& Teller paper \index{de Hoffmann-Teller}ignited an avalanche of theoretical investigations of magnetohdrodynamic shocks \citep[and others]{Helfer1953,Lust1953,Lust1955,Marshall1955,Syrovatskii1957,Shafranov1957,Vedenov1958a,Vedenov1958b,Davis1958,Gardner1958,Syrovatskii1959,Ericson1959,Ludford1959,Montgomery1959,Kontorovich1959, Liubarskii1959,Colgate1959,Sagdeev1960,Sagdeev1962a,Sagdeev1962b,Vedenov1961,Fishman1960,Germain1960,Kulikovskii1961,Morawetz1961}. \cite{Friedrichs1954}  had realized already early on that the method of characteristics could be modified in a way that makes it as well applicable to magnetohydrodynamic shocks. This allowed for the formal construction of the  geometrical shapes of magnetohydrodynamic shocks developing in front of  given obstacles of arbitrary profile.\index{L\"ust, Reimar}\index{Davis, Leverett} \index{Gardner, Charles W.}\index{Montgomery, David. C.}\index{Friedrichs, K. O.}\index{Courant, Richard}\index{Kontorovich, V. M.}\index{Marshall, W.}

Production of collisionless shocks in the laboratory encountered more severe problems as the dimensions of the devices were small, temperatures comparably low and densities high such that collisional effects could hardly be suppressed. Nevertheless, first preliminary experimental results on various aspects of the structure of nearly collisionless shocks were reported by \cite{Bazer1958}, \cite{Patrick1959}, \cite{Wilcox1960,Wilcox1961}, \cite{Auer1961,Auer1962},  \cite{Keck1962}, \cite{Camac1962}, \cite{Fishman1962}, \cite{Brennan1963} and others. The first successful production of collisionless shocks in laboratory experiments \citep{Kurtmullaev1965,Paul1965,Eselevich1971} found that the collisionless shocks investigated were highly nonstationary and exhibited complicated substructuring depending on Mach number, including strongly heated electrons \citep{Paul1967} and electric potential jumps extending over $\sim 100\lambda_D$, many Debye lengths $\lambda_D$ inside the shock transition \citep{Eselevich1982}. The existence of heated electrons and electric potentials already led \cite{Paul1967} to speculate that, at higher Mach numbers, shocks could be {\it in principle} non-stationary. Further laboratory studies by \cite{Morse1972a} and \cite{Morse1972b} with the facilities available at that time seemed to confirm this conjecture.\index{Debye, Peter}

\begin{figure}[t!]
\hspace{0.0cm}\centerline{\includegraphics[width=0.5\textwidth,clip=]{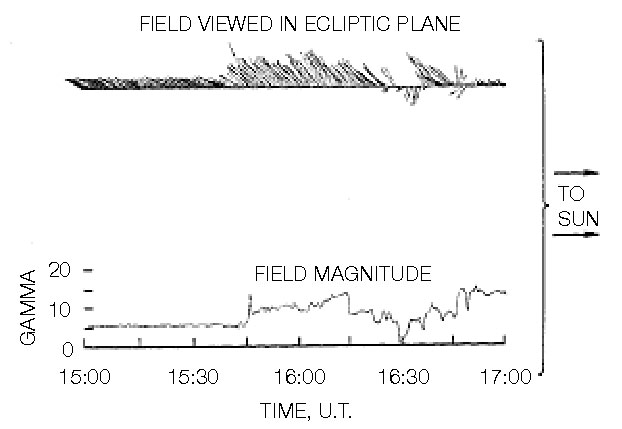} }
\caption[1]
{\footnotesize The original magnetic field recordings by Mariner II of October 7, 1962. The passage of Earth's bow shock wave occurs at 15:46 UT and is shown by the spiky increase in the magnetic field followed by a period of grossly enhanced magnetic field strength \citep[after][]{Sonett1963}. The upper part shows the direction of the magnetic field vector in the ecliptic plane and its sudden change across the bow shock.}\label{Sonett-f1}
\end{figure}
However, the first indisputable proof of the real existence of collisionless shocks in natural plasmas came from spacecraft observations when the Mariner (see Figure \ref{Sonett-f1}) and IMP satellites passed the bow shock standing upstream of Earth's magnetosphere in the solar wind \citep{Sonett1963,Ness1964}. That such shocks should occasionally exist in the solar wind had already been suggested about a decade earlier by \cite{Gold1955} who \index{Gold, Thomas}concluded that the sharp sudden commencement rise in the geomagnetic field initiating magnetic storms on Earth not only implied an impinging interplanetary plasma stream -- as had been proposed twenty years earlier by \cite{Chapman1930,Chapman1931} --  but \index{Chapman, Sidney}\index{Chapman-Ferraro model}required a very high velocity ${\cal M}>1$ solar wind stream that was able to (temporarily) create a bow shock in front of the Earth when interacting with the Earth's dipolar geomagnetic field. \cite{Zhigulev1960} and somewhat later \cite{Axford1962} and \cite{Kellogg1962}, simultaneously picking up this idea, suggested that this shock wave in front of Earth's dipole field should in fact represent a {\it stationary bow shock}  that was standing in the solar wind being at rest in Earth's reference frame.

The above mentioned seminal spacecraft observations {\it in situ} the solar wind by  \cite{Sonett1963} and \cite{Ness1964} unambiguously confirmed these claims demonstrating that the super-magnetosonic solar wind stream had traversed a thin discontinuity surface downstream of which it had entered a plasma that was in another highly disturbed irreversibly turbulent thermodynamic state. The transition was monitored mainly in the magnetic field that changed abruptly from a relatively steady solar wind value around $B_{\rm sw}\simeq$\,5-10 nT to a high downstream value that was fluctuating around an average of $\langle B\rangle\simeq$\,15-30 nT and had rotated its direction by a large angle. This shock transition surface could not be in local thermodynamic equilibrium. It moreover turned out to have thickness of the order of $\sim$ few 100 km being comparable to the gyro-radius of an incoming solar wind proton and thus many orders of magnitude less than the Coulomb mean free path. For a measured solar wind plasma particle density of  $n\sim 10\,\,{\rm cm}^{-3}$ and an electron temperature of $T_e\sim 30$\,eV the Coulomb mean free path amounts roughly to about $\lambda_{\rm C}\sim 5$ AU, vastly larger than the dimension of the entire Earth's magnetosphere which has an estimated linear extension in the anti-sunward direction of $\lesssim 1000\,{\rm R_ E} \approx 3\times 10^{-3}$\,AU. Thus Earth's bow shock represents a truly collision-free shock transition. 

This realization of the extreme sharpness of a collisionless shock like the Earth's bow shock immediately posed a serious problem for the magnetohydrodynamic description of collisionless shocks. In collisionless magnetohydrodynamics there is no known dissipation mechanism that could lead to the observed extremely short transition scales $\Delta\sim r_{ci}$ in high Mach number flows which are comparable to the ion gyro-radius $r_{ci}$. Magnetohydrodynamics neglects any differences in the properties of electrons and ions and thus barely covers the very physics of shocks on the observed scales. In its frame shocks are considered as infinitely narrow discontinuities, narrower than the magnetohydrodynamic flow scales $L\gg \Delta\gg \lambda_d$; on the other hand, these discontinuities must physically be much wider than the dissipation scale $\lambda_d$ with all the \index{scales!dissipation}physics going on inside the shock transition. This implies that the conditions derived from collisionless magnetohydrodynamics just hold far upstream and far downstream of the shock transition, far outside the region where the shock interactions are going on. In describing shock waves collisionless magnetohydrodynamics must be used in an asymptotic sense as providing the remote boundary conditions on the shock transition. One must look for processes different from magnetohydrodynamics in order to reach at a description of the processes leading to shock formation and shock dynamics and the structure of the shock transition. In fact, viewed from the magnetohydrodynamic single-fluid viewpoint the shock should not be restricted to the steep shock front, it rather includes the entire shock transition region from outside the foreshock across the shock front down to the boundary layer at the surface of the obstacle. And this even holds as well for the two-fluid shock theory that distinguishes between the behaviour of electrons and ions in the plasma fluid. 

\subsection{Early collisionless shock investigations}
\noindent Evolutionary models of magnetohydrodynamic shocks \citep{Kantrowitz1966} were based on the assumption of the dispersive evolution of one of the three magnetohydrodynamic wave modes,  the compressive fast and slow magnetosonic modes and the incompressible intermediate or Alfv\'en wave. It is clear that in a non-dissipative medium the evolution of a shock wave must be due to the nonlinear evolution of a dispersive wave disturbance. In those modes where the shorter wavelength waves have higher phase velocity the faster short wavelength waves will overcome the slower long wavelength waves and cause steepening of the wave. If there is neither dissipation nor dispersion, the wave will start breaking. Wave steepening can be balanced by dispersion, which leads to the formation of large amplitude isolated solitary wave packets. When, in addition, the steepening causes shortening of the wavelength until the extension of the wave packet in real space becomes comparable to the internal dissipation scale, a shock ramp may form out of the solitary structure. This ramp separates the compressed downstream state from the upstream state.  Such processes have been proposed to take place in an early seminal review paper by  \cite{Sagdeev1966} on the \index{Sagdeev, Raoul Z.} collective processes involved in the evolution of collisionless electrostatic shocks that was based on the earlier work of this author \citep{Sagdeev1960, Sagdeev1962a, Sagdeev1962b, Moiseev1963a, Moiseev1963b}. The ideas made public in that paper were of fundamental importance for two decennia of collisionless shock research that was based on the nonlinear evolution of dispersive waves. In particular the insight into the microscopic physical processes taking place in collisionless shock formation and the introduction of the equivalent potential  method (later called `Sagdeev potential' method) clarified \index{Sagdeev pseudo-potential}many open points and determined the direction of future shock research. \index{Kantrowitz, Alfred}\index{Petschek, H. E.}

However, even the first {\it in situ} observations identified the collisionless shocks in the solar wind like Earth's bow shock as magnetized shocks. It is the magnetic field which determines many of their properties. On the other hand, the magnetic field complicates the problem substantially by multiplying the number of possible plasma modes, differentiating between electron and ion dynamics, and increasing the possibilities of nonlinear interactions. On the other hand, the presence of a magnetic field introduces some rigidity and ordering into the particle dynamics by assigning adiabatically invariant magnetic moments $\mu=T_\perp/|{\bf B}|$ to each particle of mass $m$, electric charge $q$, and energy $T_\perp=\frac{1}{2}mv_\perp^2$ perpendicular to the magnetic field ${\bf B}$.

Because of the obvious importance of dispersive and dissipative effects in shock formation, for more than one decade the theoretical efforts concentrated on the investigation of the dispersive properties of the various plasma modes, in particular on two-fluid and kinetic modes \citep[for example]{Gary1971} and on the generation of anomalous collision frequencies \citep[for example]{Krall1969,Krall1971,Biskamp1971} in hot plasma even though it was realized already by \cite{Marshall1955} that high-Mach number shocks cannot be sustained alone by purely resistive dissipation like anomalous resistivity and viscosity. This does not  mean that the investigation of anomalous resistivity and viscosity by itself would make no sense. At the contrary, it was realized very early that wave-particle interactions replace binary collisions in collisionless plasmas thereby generation anomalous friction which manifests itself in anomalous transport coefficients, and much effort was invested into determining these coefficients \citep[e.g.][and others]{Vekshtein1970,Bekshtein1970,Liewer1973,Sagdeev1979}. These anomalous transport coefficients do in fact apply to low Mach number shocks. However, when the Mach number exceeds a certain - surprisingly low and angular dependent - critical limit the anomalous resistive or viscous time scales that depend on the growth rates of instabilities become too long in order to generate the required dissipation, heating and increase in entropy fast enough for maintaining a quasi-stationary shock. The critical Mach number estimated  by \cite{Marshall1955} was ${\cal M}_{\rm crit}=2.76$ for a perpendicular shock. This follows from the condition that ${\cal M}\leq 1$ right downstream of the shock implying that the maximum downstream flow speed $V_2=c_{s2}$ should just equal the downstream sound speed in the shock-heated flow \citep{Coroniti1970a}. At higher Mach numbers the solution that nature finds is that the shock ramp specularly reflects back upstream an ever increasing portion of the incoming plasma corresponding to the fraction of particles whose excess motional energy the shock is unable to convert into heat. 

Shock reflection had first been suggested and inferred as an important mechanism for shock dissipation by \cite{Sagdeev1966}. In a magnetized perpendicular shock like part of Earth's bow shock the shock-reflected ions generate a magnetic foot in front of the shock ramp, as has been realized by \cite{Woods1969,Woods1971}. This foot is the magnetic field of the current carried by the reflected ion stream that drifts along the shock ramp in the respective crossed magnetic field and magnetic field gradients. It is immediately clear that the efficiency of reflection must depend on the angle $\thetabn=\cos^{-1}({\bf B_1},{\bf n})$ between the shock ramp, represented by the shock normal vector ${\bf n}$, and the upstream magnetic field ${\bf B_1}$. This angle allows to distinguish between perpendicular $\thetabn=90^\circ$ and parallel $\thetabn=0^\circ$ magnetized shocks. \index{shocks!electrostatic}\index{shocks!magnetized}In the ideal reflection of a particle it is its flow velocity component $V_n$ parallel to the shock normal that is inverted. Since the particles are tied to the magnetic field by gyration, they return \index{shocks!shock-normal angle $\thetabn$}upstream with  (at most) the projection $v_\|=-V_n\cos\,\thetabn$ of this component onto the upstream magnetic field. In a perpendicular shock $v_\|=0$ and the reflected particles do not really return into the upstream flow but perform an orbit of half an gyro-circle upstream extension around the magnetic field ${\bf B_1}$.  In the intermediate domain of quasi-perpendicular $\thetabn>45^\circ$ and quasi-parallel $\thetabn<45^\circ$ shocks particles leave the shock upstream along the magnetic field ${\bf B_1}$, but the efficiency of reflection decreases with decreasing angle $\thetabn$ such that, theoretically, for parallel shocks no particles are reflected at all. 

This reflection process depends on the shock potential, height of magnetic shock ramp, shock width and plasma wave spectrum in the shock. Because of the complexity of the equations involved its investigation requires extensive numerical calculations. Such numerical simulations using different plasma models have been initiated in the early seventies \citep{Forslund1970,Biskamp1972a,Biskamp1972b} and have since become the main theoretical instrumentation in the investigation of collisionless shocks accompanying and completing the wealth of data obtained from {\it in situ} measurements of shocks in interplanetary space and from remote observations using radio emissions from travelling interplanetary shocks or the entire electromagnetic spectrum as is believed to be emitted from astrophysical shocks from the infrared through optical, radio and X-rays up to gamma rays which have been detected from highly relativistic shocks in astrophysical jets.  
\vspace{-0.3cm}
\subsection{Three decades  of exploration: Theory and observation}
\noindent During the following three decades many facets of the behaviour of collisionless shocks could be clarified with the help of these observations and support from the first numerical simulations. For the initial period the achievements have been summarized at different stages by \cite{Anderson1963}, who reviewed the then known magnetohydrodynamic shock wave theory which still was not aware of the simulation possibilities opened up by the coming availability of powerful computers, and in the first and much more important place by \cite{Sagdeev1966}, who gave an ingenious summary of the ideas that had been developed by him  \citep{Sagdeev1960,Sagdeev1962a,Sagdeev1962b} and his coworkers \citep{Vedenov1958a,Vedenov1958b,Vedenov1961,Kadomtsev1963,Moiseev1963a,Moiseev1963b,Galeev1963a, Galeev1963b,Kadomtsev1963,Karpman1964a,Karpman1964b,Karpman1964} on the nonlinear evolution of collisionless shocks from an initial disturbance growing out of a plasma instability. He demonstrated particularly clearly the physical ideas of dispersive shock wave formation and the onset of dissipation and advertised the method of the later-so called Sagdeev potential.This first seminal paper led the foundation for a three decade long fruitful research in nonlinear wave structures and shock waves.\footnote{Historically it is interesting that the Sagdeev potential method was in fact independently used already by \cite{Davis1958} in their treatment of one-dimensional magnetized magnetohydrodynamic shocks, who numerically calculated shock solutions but missed the deeper physical meaning of the Sagdeev potential which was elucidated in the work of Sagdeev. We also note in passing that the method of transforming an arbitrary second order one-dimensional differential equation into the equation of a particle moving in an equivalent potential well had been used by Kirchhoff and Kohlrausch in the second half of the nineteenth century in the context of solving the telegraph equation, long before it was independently rediscovered by Sagdeev.}  

This extraordinarily important step was followed by the next early period that was represented by the review papers of  \cite{Friedman1971}, \cite{Tidman1971}, \cite{Biskamp1973}, \cite{Galeev1976}, \cite{Formisano1977}, \cite{Greenstadt1979}, again \cite{Sagdeev1979} and ultimately two large review volumes edited by \cite{Stone1985} and \cite{Tsurutani1985}.  

These last two volumes in particular summarized the state of knowledge reached in the mid-eighties. This knowledge was based mainly on the first most sophisticated observations made by the ISEE 1 \& 2 spacecraft which for a couple of years regularly traversed the Earth's bow shock wave at many different positions and in addition crossed a number of interplanetary travelling shocks. Great discoveries made by these spacecraft were -- among others -- the division of the bow shock into quasi-perpendicular and quasi-parallel parts, the identification of the upstream structure of the bow shock \index{bow shock}which has been found to be divided into the undisturbed solar wind in front of the quasi-perpendicular bow shock, the foot region and the evolution of the ion distribution across the quasi-perpendicular bow shock \citep{Paschmann1982,Sckopke1983}, and the extended foreshock region upstream of the quasi-parallel part of the bow shock thereby confirming the earlier theoretical claims mentioned above that the supercritical shock should reflect part of the solar wind back upstream. 

Much effort was invested into the investigation of the foreshock which was itself found to be divided into a narrow electron foreshock and an extended ion foreshock, the former being separated sharply from the undisturbed solar wind.  In the broad ion foreshock region, on the other hand, the ion distribution functions were found evolving from a beam-like distribution at and close to the ion foreshock boundary towards a nearly isotropic diffuse distribution deep into the ion foreshock \citep{Gosling1978,Gosling1982,Paschmann1981} . 

An important observation was that the reflected ions become accelerated to about four times solar wind energy when being picked-up by the solar wind and coupling to the solar wind stream, a fact being used later to explain the acceleration of the anomalous component of cosmic rays in the heliosphere. It was moreover found that the reflected ions strongly interact with the upstream solar wind \citep{Paschmann1979} via several types of ion beam interaction \citep{Gary1981}, first observed a decade earlier in the laboratory by \cite{Phillips1972}. These instabilities were found generating broad spectra of low frequency electromagnetic modes that fill the foreshock with an intense spectrum of low frequency electromagnetic fluctuations that propagate in all directions with and against the solar wind. A professional timely review of the observations of all waves observed in the foreshock was given by \cite{Gurnett1985} and is contained in the above cited volume edited by \cite{Tsurutani1985}. Some of these lowest frequency fluctuations can even steepen and generate large amplitude low frequency or quasi-stationary wave packets in the solar wind resembling small spatially localized and travelling magnetic ramps or shocklets. When the solar wind interacts with these shocklets, the solar wind stream becomes already partially retarded long before even reaching the very shock front. In other words, the entire foreshock region is already part of the shock transition. this has for long time not been realised even though it was suggested by observations.

Another important observation concerned the evolution of the electron distribution in the vicinity of the bow shock and particularly across the quasi-perpendicular bow shock \citep{Feldman1982, Feldman1983}. It was in fact found that the electron distribution evolved from the solar wind nearly Boltzmannian plus halo distribution to make the transition to a flat-top heated electron distribution when crossing the shock ramp.  The flat top suggested that strong electron heating takes place inside the shock on a short time scale and that the shock ramp contains a stationary electric field which is partially responsible for the reflection of particles back upstream while at the same time heating the in-flowing electrons. Moreover the distributions showed that electrons were also reflected at the shock escaping in narrow beam bundles into the upstream solar wind along the tangential to the bow shock solar wind magnetic field. Upstream of the shock they were strong enough to generate plasma fluctuations around the plasma frequency and to give rise to radio emission at the harmonic of the plasma frequency. Plasma wave observations \citep{Rodriguez1975,Anderson1981} were confirmed by these observations as being excited by the shock reflected electron beams. 

The period following the comprehensive reviews by \cite{Stone1985} and \cite{Tsurutani1985} were devoted to further studies of the bow shock by other spacecraft like AMPTE, and of other interplanetary shocks accompanying solar ejection events like CMEs by Ulysses and other spacecraft and in particular by the Voyager 1 \& 2 satellites who investigated travelling shocks and corotating with the sun interacting regions in interplanetary space, heading for encounters with bow shocks of the outer planets and ultimately the heliospheric termination shock. To everyone's excitement the heliospheric termination shock was crossed in a spectacular event in December 2004 by Voyager 1 at a distance of $\sim$94 AU confirming the prediction of its existence and at the same time opening up a large number of new problems and question that had not been expected or anticipated before. Ten years after the above two reviews, \cite{Russell1995} edited a proceedings volume collecting the accumulated research of this period. And another ten years later \cite{Li2005} organized a conference reviewing the more recent achievements in shock research. 

\subsection{The numerical simulation age}
\noindent The advent of powerful computer resources in the mid-sixties completely changed the research attitudes also in shock physics research. Starting from the idea that all the physics that can be known is contained in the basic physical laws which can be represented by a set of conservation equations (below we will provide a number of such model conservation laws), the idea arose that searching the domain of solutions of these conservation laws in the parameter space prescribed by observations with the help of the new computing facilities would not only liberate one from the tedious burden of finding mathematically correct analytical solutions of these laws but would also expand the accessible domain of solutions into those directions where no analytical solutions could be found. With this philosophy in mind a new generation of researchers started developing numerical methods for solving the basic nonlinear equations of plasma physics with the help of powerful computer facilities circumventing the classical methods of solving partial differential equations and enabling attacking any nonlinear problem by sometimes straight forward brute force methods. In fact, 
collisionless shocks are particularly well suited for the application of such methods just because they are intrinsically nonlinear. 

The decades since the late sixties were thus marked by an inflation of numerical approaches, so-called computer simulations, to shock physics thereby paralleling similar developments in all fields of exact scientific research. The ever increasing capacities of the computer have been very tempting for performing simulations. However, the capacities are still suited only for very well tailored simulation problems. It is moreover clear that numerical simulations do not allow to make fundamental discoveries which go ahead of the amount of information that is already contained in the equations one is going to solve on the computer. Still this is an infinite number of problems out of which the relevant and treatable must be carefully and insightfully extracted. In addition any simulation requires the application of subsequent data analysis which closely resembles the analysis of observational data obtained in real space or laboratory experiments. Because of this second reason one often speaks more correctly of computer experiments or numerical experiments instead of numerical simulations.
\begin{figure}[t!]
\centerline{\includegraphics[width=0.9\textwidth,height=0.5\textwidth,clip=]{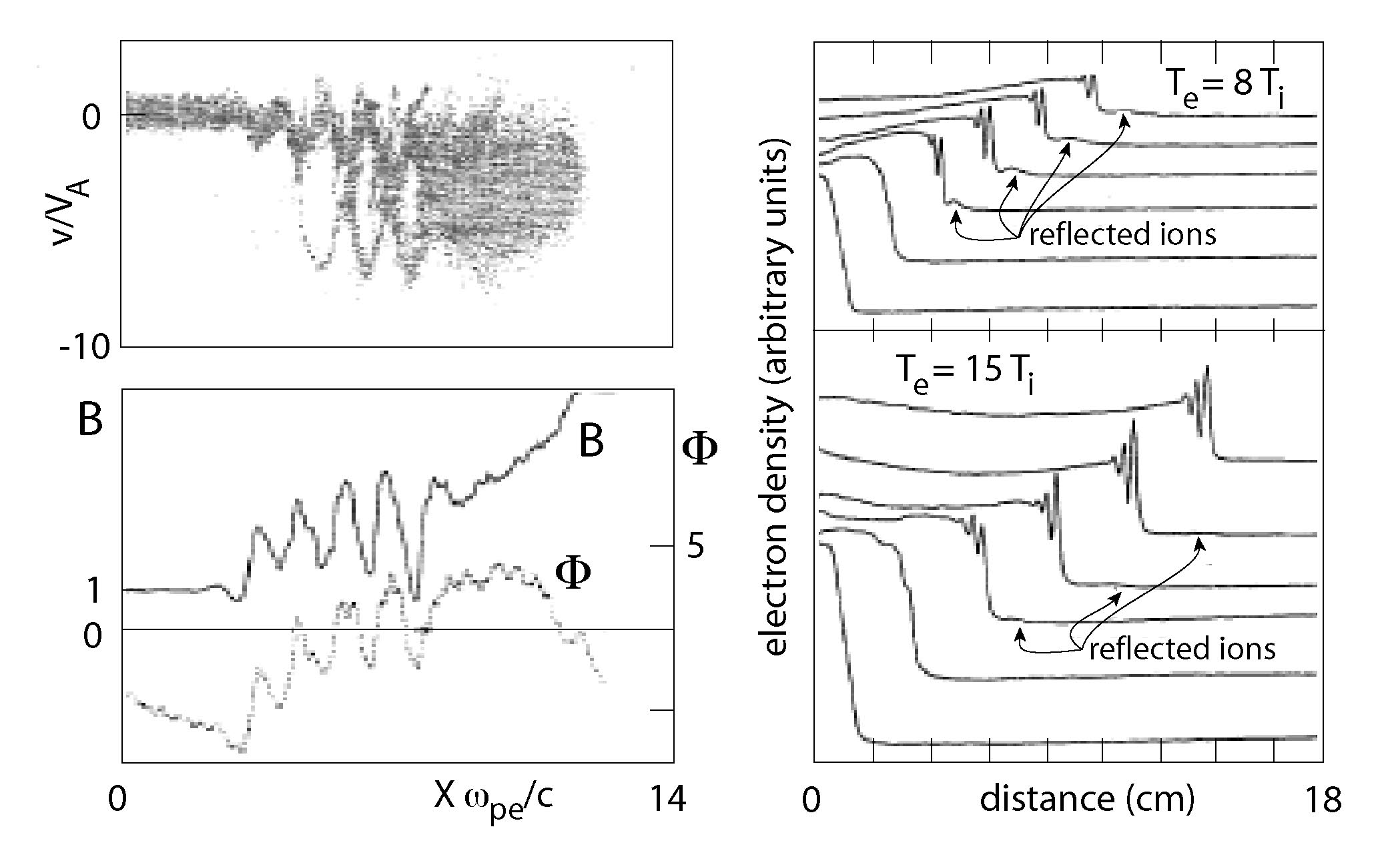} }
\caption[1]
{\footnotesize One-dimensional numerical simulations by (right) \cite{Taylor1970} and (left) \cite{Biskamp1972a} of shocks with reflected ions. In Taylor et al.'s simulation the shock ramp is shown in time-stacked profiles with time running from left corner diagonally upward. The density ramp is shown. The initial profile has been assumed as a steep ramp. Reflection of ions leads with increasing time to its oscillatory structure and generation of a foot as had been suggested by \cite{Woods1969}. Biskamp \& Welter's simulation shows the phase-space (upper panel), the evolution of reflected (negative speed) ions and heating behind the shock ramp. In the lower panel the evolution of the magnetic field with distance in magnitude and direction is shown. The field exhibits strong undulations caused by the gyrating reflected ions in the solar wind in front of the ramp. A rotation of the field angle through the shock is also detected. }\label{Biskamp-f1}
\end{figure}

There are two fundamentally different directions of simulations \citep[for a collection of methods, see, e.g.][]{Birdsall1991}. Either one solves the conservation laws which have been obtained to some approximation from the fundamental Liouville equation, or one goes right away back to the number of particles that is contained in the simulation volume and solves for each of them the Newtonian equation of motion in the self-generated fields. Both methods have been applied and have given successively converging results. Both approaches have their advantages and their pitfalls. Which is to be applied depends on the problem which one wants to solve. For instance, in order to determine the global shape of the Earth's bow shock it makes no sense to refer to the full particle approach; a fluid approach is good enough here. On the other hand if one wants to infer about the reflection of particles from the shock in some particular position on the smaller scale, a full particle code or also a Vlasov code would possibly be appropriate.  
 
\begin{figure}[t!]
\hspace{0.0cm}\centerline{\includegraphics[width=0.95\textwidth,height=10cm,clip=]{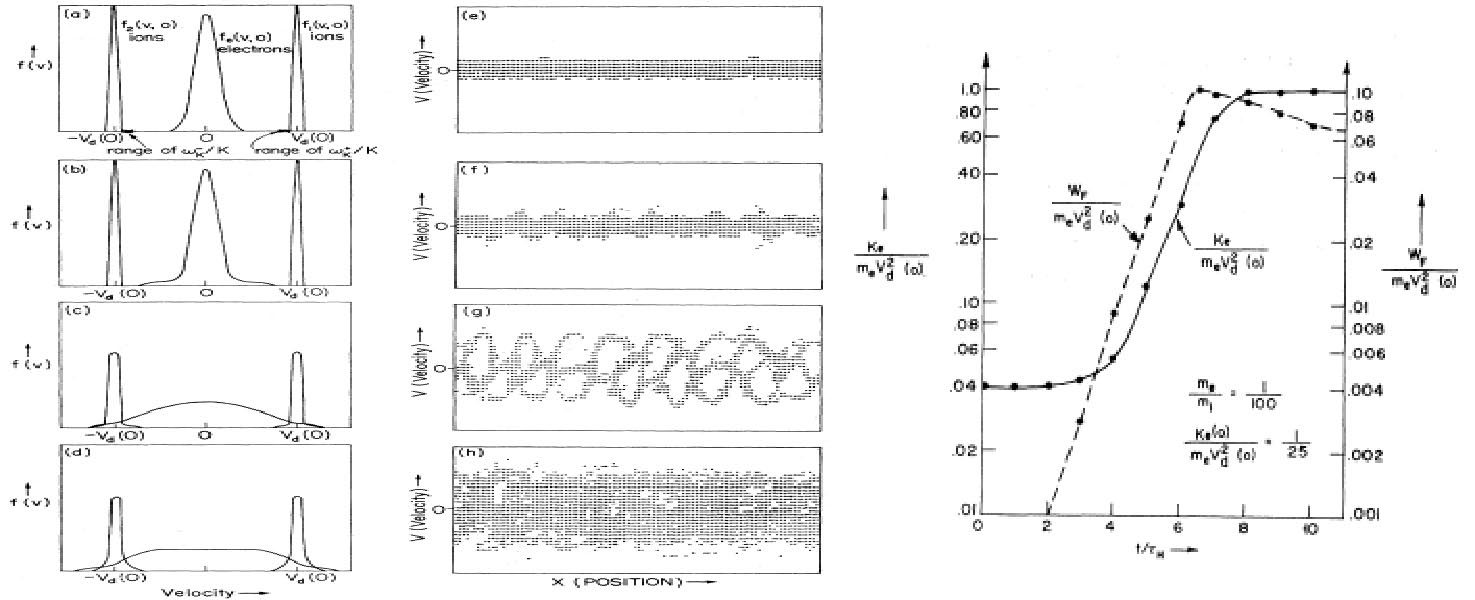} }
\caption[1]
{\footnotesize A one-dimensional purely electrostatic particle simulation of shock plasma heating as observed by \cite{Davidson1970}. The (unrealistic model) simulation in the shock reference frame consists of a thermal electron background and two counter-streaming ion beams (left column) of exactly same temperature and density modelling the inflow of plasma (forward ion beam) and reflected ions (backward beam). The electrons are assumed hotter than ions neglecting their bulk motion but cold enough for allowing the Buneman instability \citep{Buneman1959} to develop which requires a strong shock. The left panel shows the evolution of the distribution functions: broadening of all distributions in particular of electrons indicating the heating. The middle column is the electron phase space representation showing the disturbance of electron orbits, increase of phase space volume and final thermalization. Probably the most interesting finding is in the third phase space panels which shows that as an intermediary state the electron orbits show formation of voids in phase space. These holes contain strong local electric fields but were not recognised as being of importance at this time. They were independently discovered theoretically only a few years later \citep{Schamel1972,Schamel1973,Schamel1979,Dupree1975,Dupree1982,Dupree1983,Dupree1986,Berman1985}. The right panel shows the evolution of electron kinetic and electric field energy during the evolution. }\label{Davidson-f1}
\end{figure}

In collisionless shock physics computer experiments have proven very valuable since the beginning. In order to elucidate the internal physical structure of shocks one is, however, directed to particle codes rather than fluid codes. First electrostatic simulations of one-dimensional collisionless shocks have been performed with very small particle numbers and in very small simulation boxes (spatio-temporal boxes with one axis the space coordinate, the other axis time) by \cite{Dawson1968}, \cite{Forslund1970} and \cite{Davidson1970}. They observed the expected strong plasma heating (see Figure \ref{Davidson-f1}. Slightly later \cite{Biskamp1972a,Biskamp1972b} in similar one-dimensional simulations found the first indication for reflection of particles in shock formation.  With larger simulation boxes a decade later the same authors \citep{Biskamp1982} confirmed that  in the simulations high Mach number shocks indeed reflected ions back upstream when the Mach number exceeded a certain critical value. 

These first full particle simulations were overseeded in the eighties by hybrid simulations, which brought with them a big qualitative and even quantitative step ahead in the understanding of collisionless shocks \citep{Leroy1982,Leroy1983,Leroy1984,Leroy1984a,Scholer1990,Scholer1990a} a method where the ions are treated as particles while the electrons are taken as a fluid. These were all one-dimensional simulations where an ion beam was allowed to hit a solid wall until being reflected there. Electrons were treated as a massless isothermal fluid. The perpendicular shock evolved in the interaction of the incoming and reflected ion beams and in the frame of the wall moved upstream at a certain measurable speed. 

The first of these simulations were of subcritical perpendicular shocks showing the effect of shock steepening and  included some numerical resistivity the effect of which could be inferred about. Increasing the Mach number then showed the transition to shock ion reflection and formation of the foot region in front of the perpendicular shock. The later simulations then included oblique magnetic fields and started investigating the reflection process in dependence on the shock normal angle and inferring about the particle reflection and acceleration processes. These hybrid simulations were capable of reproducing various properties of collisionless shocks like the ion-foot formation in perpendicular and quasi-perpendicular shocks and ion reflection and shock reformation in quasi-parallel shocks. They also allow to distinguish between subcritical and supercritical shocks as well as inferring about the various low-frequency waves which are excited in the ion foreshock by the super-critical shock-reflected ion distribution. Clearly, since the electrons act only passively as a neutralizing fluid their dynamics is neglected in these simulations. Nevertheless a number of important properties of collisionless shocks was recovered in these simulations of which the above mentioned foot and the shock reformation are among the most important. In addition it was found that contrary to the pure fluid theory the supercritical shocks turned out to possess a self-generated non-coplanar magnetic field component. And it was also found that the shock ramp itself was nonstationary and behaved stationary only in the average over scales long compared with the reformation time scale and extended along the shock front. 

Clearly hybrid simulations can be used to describe this meso-scale structure of shocks, but most of the interesting shock physics takes place on shorter and smaller scales when electrons must be treated as particles as well. 

Currently the investigation of collisionless shocks has achieved a certain stage of saturation. Analytical theories have basically been exhausted for the last fifteen years. The complexity of shock structure has set a fairly strict bound to it unless completely new mathematical methods will be brought up allowing for taking this complexity into account without further increasing the lacking mathematical simplicity and transparency. 

Measurements {\it in situ} have as well been pushed to their bounds. Improved instrumentation and the increase in on-board data storage capabilities and speed of on-board data analysis and data reduction provided an already available enormous data pool which is by far not yet exhausted. However, most of the data obtained are from single spacecraft observations and thus suffer from being single point measurements such that it is difficult or even impossible to distinguish between spatial effects and temporal evolution. 

On the other hand, the multi-spacecraft observations like combinations of different spacecraft suffer from the accidental character of their connectedness with the bow shock or other shocks in interplanetary space. Specially designed multi-spacecraft missions like Cluster suffer from the inflexibility of tuning the sub-spacecraft separations. Some of these pitfalls can be overcome by sophisticated data analysis methods \citep{Paschmann2000} but the progress is not as overwhelming as one would hope. 
Moreover, between design and launch of such multi-spacecraft lies roughly one decade such that the instrumentation cannot keep track with the development in technology which is available when the spacecraft is put into orbit. 

Finally, even though the computing capabilities increase almost exponentially, the currently available computers still cannot manage to solve the orbits of the vast numbers of particles in a realistic volume in space and phase space. For instance the spatial volume at the perpendicular bow shock wave of Earth amounts to roughly $50\,{\rm R_E}^3\approx 2 \times 10^{19}\,{\rm m}^3$ which for an average density of $n\sim 10^7\,{\rm m}^3$ implies just $N\sim 10^{26}$ protons plus the same number of electrons in the volume and passing through the shock at a speed of a few 100\,km/s. The orbits of such numbers of particles in their self-excited and external fields cannot be calculated even by the most powerful computers. 

Hence, in particle codes particles must be grouped together into macro-particles neglecting their short distance interactions and short distance fields. As long as this grouping affects small particle numbers its effects on the final results are certainly negligible. However, since computers so far can manage to solve only up to the order of few times $\sim\!10^9$ particle orbits simultaneously, investigation of the full perpendicular shock with particle codes requires a macro-particle size of the order of $\sim\!10^{17}$ individual particles which is not small anymore but forms super-clusters of particles with all their unknown internal and external dynamics. Since at a density of $n\!\sim\! 10^7\,{\rm m}^{-3}$ each particle occupies an elementary volume of $n^{-1}\sim 10^{-7}\,{\rm m}^3$, each of these macro-particles occupies a volume of $\sim\!10^{10}\,{\rm m}^3$ or a plasma blob of  $10\,{\rm km}^3$ volume. 

For comparison, in the solar wind the Debye length is $\lambda_D\sim 23$\,m; the {\it real} Debye volume is thus six orders of magnitude smaller than the macro-particle volume, negligibly small compared with the macro-particle volume and cannot be resolved in three-dimensional full particle simulations. Any resolvable linear spatial scale must be larger than $\ell_{\rm MP}\sim 5$\,km. The electron gyro-radius is about 200\,km, the electron inertial length about 1700\,km such that these lengths can be resolved. Similarly, any proton scale can be resolved as well. However care must be taken when simulating problems where the electron Debye scale is involved like in radiation problems. Moreover since the scale of macro-particles is a real space scale, the validity of any one- or two-dimensional simulations in the directions of the neglected coordinates is restricted to scales $\gg 5$\,km which in the solar wind is not a severe restriction. In order to resolve the Debye scale in radiation problems the system is assumed to be homogeneous over $\sim 500-1000$\,km in two-dimensional and $50,000-100,000$\,km in one-dimensional simulations in the directions on which coordinates the simulation is independent. At the same time, however, in the remaining simulation direction the resolution can be made sufficiently high enough to resolve the relevant times and lengths.  

However, the main problem is not the resolution itself. It is the neglect of the interactions between the particles constituting the macro-particles since these are dynamical systems which continuously exchange particles, momentum, and energy among themselves which may or may not introduce systematic errors. The limitations on the simulation results are very difficult to estimate, however, and in most cases are believed to be unimportant. Conservation of macro-particle number and total energy are usually good measures of this validity of the simulations which are usually stopped when these conservations becomes violated in the course of the runs, and the results are taken valid for simulation times only where the energy and particle numbers were conserved within a few per cent. However, even in this case it must be stressed that from simulations it can only be inferred what is already contained in the equations. They are not suited for making any real physical discoveries while they can make predictions that can be checked against observation thereby validating or falsifying theory.

\subsection{Scope of this Paper} 
\noindent The present Chapter is the result of an attempt to give a more systematic overview of our current knowledge of collisionless shocks based on the available in situ observations, measurements from space as well as the results obtained in the enormous number of efforts in simulations and analytical theory invested during the past decades. It will turn out that during the past few decades a more coherent view of the structure of a collisionless shock has indeed been achieved. To round up the current state of the art in collisionless shock physics and lay the grounds for further research on this subject it seems reasonable if only preliminary to attempt building a first coherent and comprehensive picture of the physics of collisionless shocks.

\section{When Are Shocks?}
\noindent At a first glance the notion of a collisionless shock seems nonsensical \citep[for a popular review see][]{Sagdeev1991}. This is also the first reaction one encounters when talking to either laymen or even physicist working in a different area of physics. Usually one earns a forgiving smile from either of them. Indeed, one intuitively imagines that a shock requires that something is shocked by collisions. However, Nature is not always organized the way one naively believes. Though ordinary gasdynamic shocks or shocks in condensed matter are indeed shocks which cannot be thought of when ignoring the high collisionality, i.e. ignoring the short range forces involved in binary collisions when the particles (or solid objects) literally touch and hit each other, collisionless systems like high temperature dilute plasmas take advantage of nonlocal, non-binary, `anomalous collisions' between\index{collisions!anomalous} particles and the existing external and selfconsistently generated electromagnetic fields. These `anomalous collisions' are in fact long range {\it collective interactions} between \index{interactions}\index{process!collective}groups of particles and fields; they lead to correlations between these groups such that the particles do not anymore behave like free ballistically moving particles. Because of the correlations the particles together with the fields organize themselves to form structures. The possibility of such structure formation in completely collision-free plasma has been an extraordinarily important insight. It leads to close interaction and the generation of irreversible dissipation processes in plasma that result in heating, acceleration of groups of particles, generation of entropy and emission of radiation, which all are different forms of energy distribution in and energy loss from the plasma. It is the very meaning of these collective processes to make the otherwise completely dissipation-free plasma capable of returning to a thermodynamic state of destruction of the available free energy and transforming it into heat and other less valuable forms of energy. One of these structures are shock waves.

Collisionless shock waves form when a large obstacle is put into a plasma flow that is either super-Alfv\'enic or super-magnetosonic in the frame of the obstacle.  Super-Alfv\'enic flows  have Mach numbers ${\cal M}_A=V/V_A>1$, where ${\bf V}_A= {\bf B}/\sqrt{\mu_0m_iN}$ is the Alfv\'en velocity in the magnetized plasma of density $N$ and magnetic field ${\bf B}$, and $m_i$ is the ion mass. Correspondingly, super-magnetosonic flows have Mach numbers ${\cal M}\equiv {\cal M}_{ms}=V/c_{ms}>1$, where $c_{ms}^2= V_A^2+c_s^2$ is the square of the magnetosonic speed, and $c_s^2=\partial P/\partial \rho$ is the square of the ordinary sound speed, with $P$ the isotropic pressure, and $\rho= m_iN$ the mass density. In plasma of respective electron and \index{waves!magnetosonic}ion temperatures $T_e$ and $T_i$ (in energy units) and adiabatic indices $\gamma_{e,i}$ for electrons and ions the latter is with sufficient accuracy given by $c_s^2= (\gamma_eT_e+\gamma_iT_i)/m_i$. We should, however, note that this discussion is based on fluid considerations, and even then it is valid in this form only for the so-called fast magnetosonic mode perpendicular to the magnetic field ${\bf B}$. \index{waves!Alfv\'en}The magnetosonic speed depends on the wave propagation angle $\theta$ with respect to the magnetic field. Its general fluid expression
\begin{equation}\label{chap1-eq-cms}
c_{ms}^2(\theta)= c_{ms}^2\pm\left[(V_A^2-c_s^2)^2+4V_A^2c_s^2\sin^2\theta\right]^\frac{1}{2}
\end{equation}
where $c_{ms}$ is the angle-independent expression given above, shows that the magnetosonic velocity contains two branches, $c_{ms}^+$ related to the {\it fast} magnetosonic wave mode with the positive sign in front of the root, and $c_{ms}^-$ related to the {\it slow} magnetosonic wave mode with negative sign in front of the root. In addition there is the Alfv\'en wave with speed $V_A$ that is independent of the sound velocity $c_s$. Of these three modes only the fast mode propagates perpendicular to the magnetic field. However in all other directions all three modes can exist, and therefore each of them may form a shock if only the flow speed exceeds its velocity. Hence there can, in principle, exist three different kinds of shocks, the {\it fast}, {\it slow}, and {\it Alfv\'enic} (or {\it  intermediate}, because its speed is intermediate between the fast and slow waves) shocks, respectively.

\begin{figure}[t!]
\hspace{0.0cm}\centerline{\includegraphics[width=0.975\textwidth,clip=]{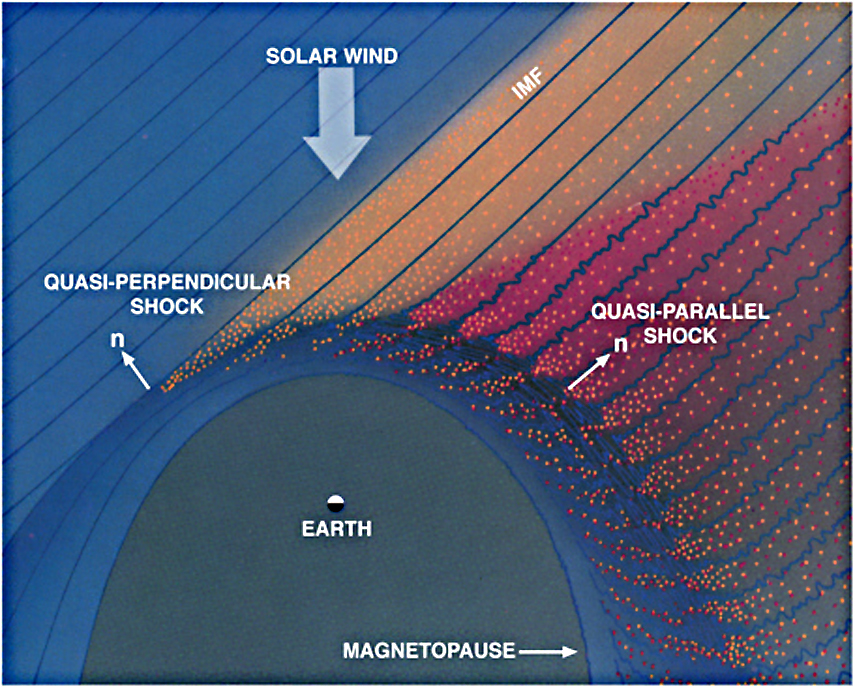} }
\caption[1]
{\footnotesize A two-dimensional schematic view on Earth's steady-state bow shock in front of the blunt magnetosphere \citep[from][]{Tsurutani1985} which forms when the supersonic solar wind streams against the dipolar geomagnetic field. The bow shock is the diffuse hyperbolically shaped region standing at a distance in front of the magnetopause. The inclined blue lines simulate the solar wind magnetic field (interplanetary magnetic field IMF). In this figure the lie in the plane. The direction of the shock normal is indicated at two positions. Where it points perpendicular to the solar wind magnetic field the character of the bow shock is perpendicular. In the vicinity of this point where the solar wind magnetic field is tangent to the bow shock the shock behaves quasi-perpendicularly. When the shock is aligned with or against the solar wind magnetic field the bow shock behaves quasi-parallel. Quasi-perpendicular shocks are magnetically quiet compared to quasi-parallel shocks. This is indicated here by the gradually increasing oscillatory behaviour of the magnetic field when passing along the shock from the quasi-perpendicular part into the quasi-parallel part. Correspondingly, the behaviour of the plasma downstream of the shock is strongly disturbed behind the quasi-perpendicular shock. Finally, when the shock is super-critical, as is  the case for the bow shock, electrons and ions are reflected from it. Reflection is strongest at the quasi-perpendicular shock but particles can escape upstream only along the magnetic field. Hence the upstream region is divided into an electron (yellow) and an ion foreshock accounting for the faster escape speeds of electrons than ions.}\label{Tsurutani-f1}
\end{figure}

Collisionless shocks are macroscopic phenomena in which very many particles are involved. The requirement on the obstacle is that its diameter $D$ in the two directions perpendicular to the flow must be very large compared to the intrinsic scales of the flow while at the same time being much less than the collisional mean free path. The largest intrinsic scale of a magnetized plasma is the ion gyro-radius $r_{ci}= V_{i\perp}/\omega_{ci}$, where $V_{i\perp}$ is the ion flow velocity perpendicular to ${\bf B}$, and $\omega_{ci}=eB/m_i$ is the ion cyclotron angular frequency. Hence with the Coulomb mean free path $\lambda_{\rm C}=(N\sigma_{\rm C})^{-1}$ defined earlier the above condition is simply that
\begin{equation}
r_{ci}\ll D \ll \lambda_{\rm C}
\end{equation}
The shock forms an extended surface that is bent around the obstacle. An example of the form of such a shock is drawn in Figure \ref{Tsurutani-f1} showing (among various other of its properties) the average shape of the Earth's bow shock in an ecliptic cross section that is caused in the interaction of the solar wind with the -- approximately -- dipolar geomagnetic field.

The radius of curvature $R_c\gg \Delta$ of the shock perpendicular to the flow will always be much larger than the width $\Delta\sim r_{ci}$ of the shock in the direction of flow, the latter being of the order of a few ion gyro-radii only. Collisionless shocks can in good approximation be considered as thin, locally flat surfaces of width $\Delta$ and outer shock normal ${\bf n}$. Given the function of the shock surface $F_S({\bf r})=r$, where $r$ is the radius vector from the (arbitrary) centre to one point on the shock surface, the shock normal is defined as
\begin{equation}\label{chap1-eq-shocknormal}
{\bf n}({\bf r})=-\frac{\nabla F({\bf r})}{|\nabla F({\bf r})|}
\end{equation}
the negative sign telling that it points outward (the shock being seen from the inside as a concave surface), the gradient accounting for the direction of strongest variation, and the normal is normalized to the gradient since it is a unit vector that satisfies the condition ${\bf n}\cdot{\bf n}=1$.

The important consequence of the above scaling is that, locally, of all spatial derivatives $\partial/\partial x\sim\partial/\partial y\ll \partial/\partial n$  only the derivative across the shock front counts. The gradient operator $\nabla$ thus reduces to the derivative in the direction opposite to the local shock normal ${\bf n}$ or, with coordinate $n$,
\begin{equation}\label{chap1-eq-nabla}
\nabla \simeq -{\bf n}\frac{\partial}{\partial n} \sim -{\bf n} \frac{[\cdots]}{\Delta}
\end{equation}
where the brackets $[\cdots]= (\cdots)_2-(\cdots)_1$ stand for the difference of the values of the quantity under consideration downstream in region 2 behind the shock minus the value upstream  in region 1 in front of the shock. Formal theory is extensively making use of this fact. Clearly, however, it looses its importance in full particle simulations and in particular in simulations in more than one dimension. In those simulations the shock turns -- to a certain degree trivially -- out to be nonstationary, i.e. even in one-dimensional simulations the shock steepness and thus the local shape of the shock vary with time. 

Rather than its shape, the most important physical property of the shock is its capability of decreasing the upstream flow from super-magnetosonic to magnetosonic or even sub-magnetosonic speed,  i.e. from ${\cal M}>1$ to ${\cal M}\leq 1$, on the extraordinarily short distance $\Delta$ of shock width. Similar as for gasdynamic shocks this implies that a substantial amount of upstream flow energy must be converted into compression of the plasma, i.e. into enhanced pressure over the narrow length $\Delta$. This is possible only if in addition the plasma is heated over the same distance. The shock thus creates a downstream region of high pressure, producing entropy, and separates it form the lower pressure upstream region. \index{shocks!width $\Delta$}From this point of view, collisionless shocks are alike to gasdynamic shocks even though there are no collisions between particles. Other mechanisms are required bridging the lack of collisions and providing the necessary dissipation. Moreover, being a permeable boundary between two regions of different temperatures and pressures, the shock is not in thermal equilibrium. Thermal non-equilibria in closed systems cannot survive, however. They have the tendency of evolving towards thermal equilibrium until it is reached. It is therefore important to realize that a shock cannot be stationary; in order to be maintained over long periods it must be continuously reformed. This is indeed the case with all the collisionless shocks observed  in the heliosphere. Collisionless shocks as quasi-stationary non-transient phenomena occur only in open systems like in the solar wind-magnetosphere interaction and are continuously reformed on the expense of the plasma inflow.

\section{Types of Collisionless Shocks}
\noindent When speaking about shock waves the implication is that like an ordinary wave the shock propagates on the background plasma. This propagation is not obvious when looking for instance at Earth's bow shock which, \index{bow shock}when neglecting its irregular change of position, is in the long term average about stationary, standing in front of the magnetosphere in the solar wind. However, the more correct view is when looking at the shock in the frame of the moving solar wind. Seen from there the shock is propagating in the solar wind upstream towards the Sun at an approximate velocity that is comparable to the solar wind speed. In the solar wind frame the shock appears as a compressive upstream moving wave front or wave ramp like a tsunami in the ocean following a seaquake. Later we will illuminate the question how such a large amplitude shock ramp can form. In preparation of a deeper discussion we in this section provide a rough classification of the various types of shocks that can develop in a collisionless plasma.

\subsection{Electrostatic shocks}\index{shocks!electrostatic}
\noindent Shocks can be classified from various points of view. The first and simplest classification is with respect to the electrodynamic properties of shocks. Plasmas consist of electrically charged particles which under normal conditions for maintaining overall charge neutrality occur in about equal numbers per volume element. They have same number densities, $N_e=N_i=N$. When the different charges do in the average not possess different bulk velocities the plasma is free of electric currents ${\bf j}=Ne({\bf V}_i-{\bf V}_e)=0$ and, in the absence of an external magnetic field ${\bf B}_0$, the plasma is free of magnetic fields. It behaves purely electric. In this case a shock wave which occurs in the flow is called an {\it electrostatic shock}. 

In the heliosphere such electrostatic shocks are rare because most moving plasmas are magnetized. They may, however, occur under certain very special conditions even in the strongly magnetized plasmas in the auroral zones of magnetized planets and in the particle acceleration zones in the solar corona during particular flare events. In these cases electrostatic shocks are strictly one-dimensional however and occur only on very small scales where they contribute to the generation of magnetic-field aligned electrostatic fields. These fields can promptly accelerate particles to energies of the order of the total macroscopic electrostatic potential drop. We will return to this problem in this book in the chapter on particle acceleration. However, this kind of shocks does not belong to the regular large-scale genuine shocks that lie at the focus of this text. Nevertheless, the theoretical and numerical investigation of electrostatic shocks was instrumental in the understanding of shock physics.

\subsection{Magnetized shocks}\index{shocks!magnetized}
\noindent The vast majority of collisionless shocks in the heliosphere -- and as well in astrophysics -- belongs to a different class of shocks known as {\it magnetized shocks} simply because the plasmas in the heliosphere is magnetized and allows for electric currents to flow across and along the magnetic field. For instance, when a magnetized super-magnetosonic moving plasma is shocked it is quite natural that the different drift motions of particles of opposite charges generated in the plasma gradient of the shock ramp causes electric drift currents to flow across the magnetic field. These currents are accompanied by proper secondary magnetic fields and in addition cause other effects like anomalous transport and plasma heating. Magnetized shocks therefore behave quite differently from electrostatic shocks and because of their abundance are much more important. 

Since the presence of the magnetic field is the first main distinction between electrostatic and magnetized shocks it is reasonable, for a first classification, to distinguish between shocks in which the upstream magnetic field is tangential to the shock surface and those shocks where the magnetic field is perpendicular to the shock surface. The reason for such a distinction is that in the former case the upstream flow velocity, ${\bf V}_1\perp {\bf B}_1$, is perpendicular to the upstream magnetic field while, in the second case, the upstream flow is parallel to the magnetic field, ${\bf V}_1\| {\bf B}_1$. This has the consequence that in the perpendicular case the magnetic term ${\bf V}_1\times{\bf B}_1\neq 0$ in the Lorentz force is finite. the magnetic field lines are convected with the flow and pile up at the shock ramp.  In the parallel case the magnetic term in the Lorentz force vanishes identically. Naively though such shocks become unmagnetized and should behave like gasdynamic shocks. This is, however, not the case as we will see in later chapters. Shocks where the flow is parallel to the field behave very different from shocks where the flow is perpendicular. 

In order to distinguish between these two types of shocks one rather refers to the shock normal defined in Eq.\,(\ref{chap1-eq-shocknormal}) which gives a precise local definition of the shock surface. Defining the shock normal angle $\thetabn$ through
\begin{equation}\label{chap1-eq-thetabn}
\tan \thetabn = {\bf n}\cdot{\bf B}_1/|{\bf B}_1|
\end{equation}
we can then distinguish {\it perpendicular shocks} with $\thetabn=\frac{1}{2}\pi$ and {\it parallel shocks} with $\thetabn=0$. In nature the two extreme cases at a large bent shock are realized only over a small portion of the entire shock surface. One therefore rather distinguishes {\it quasi-perpendicular} and {\it quasi-parallel} shocks which are operationally defined by $\frac{1}{4}\pi< \thetabn\leq \frac{1}{2}\pi$ and $0\leq \thetabn< \frac{1}{4}\pi$, respectively. In the overlap region $\frac{1}{6}\pi< \thetabn\leq \frac{1}{3}\pi$ one speaks about {\it oblique} shocks keeping, however, in mind that in contrast to the distinction between parallel and perpendicular shocks the term oblique is conventional only and is not required by physics. Oblique shocks simply have mixed parallel/perpendicular shock properties. 

Figure\,\ref{Tsurutani-f1} shows the artistic drawing of an example of a supercritical collisionless shock wave, the Earth's bow shock, located in the super-magnetosonic and magnetized solar wind exhibiting regions of all three kinds of supercritical shock waves along its bent hyperbolically shaped surface. The outer normal ${\bf n}$ to the shock has been drawn in two point, the completely perpendicular region on the left where the interplanetary magnetic field is strictly tangential to the idealised shock surface, and in the strictly parallel region where the idealised magnetic field is about parallel to the shock normal. There is a pronounced difference in both locations shown in the drawing in reference to the observations which will be discussed in depth in Chapter 7. This difference refers to the degree of distortion of the magnetic field and shock in both positions. The perpendicular shock is considerably less disturbed than the parallel shock. The yellow and red colored regions indicate the spatial domains filled with particles, electrons and protons, reflected from the supercritical shock. In the region along the shock surface between the perpendicular and parallel parts the shock changes its character gradually from quasi-perpendicular through oblique to quasi-parallel.

The magnetic field introduces another important property of the plasma, i.e. a pressure anisotropy. The pressures parallel $P_\|$ and perpendicular $P_\perp$  to the magnetic field can become different, and the pressure is in fact a tensor $\textsf{P}=P_\perp\textsf{ I}+(P_\|-P_\perp){\bf BB}/B^2$, where $\textsf{I}$ is the unit tensor. This anisotropy has the effect that the ratio $\beta=2\mu_0P/B^2$ of thermal to magnetic field pressure in general becomes anisotropic as well, with $\beta_\|= 2\mu_0P_\|/B^2\neq \beta_\perp=2\mu_0P_\perp/B^2$. Owing to this we define an anisotropy factor
\begin{equation}\label{chap1-eq-anisotropy}
A = \frac{\beta_\perp}{\beta_\|}-1\equiv  \frac{P_\perp}{P_\|}-1\equiv  \frac{T_\perp}{T_\|}-1
\end{equation}
which can be positive for an excess in perpendicular energy, negative for an excess in parallel energy, and it also can vanish when the anisotropy is vanishingly small.

\subsection{MHD shocks}\index{shocks!MHD}
\noindent The simplest and historically also the first model approach to collisionless shocks was within magnetohydrodynamic fluid theory by simply adding to the frictionless gasdynamic equations the Lorentz force
\begin{equation}
{\bf F}_L= {\bf j}\times{\bf B}
\end{equation}
with current density ${\bf j}$ defined through Ohm's law 
\begin{equation}
{\bf j}=\sigma({\bf E}+{\bf V}\times{\bf B})
\end{equation}
Under collisionless ideally conducting conditions the conductivity is $\sigma\to\infty$, and Ohm's law is replaced by the ideal MHD condition
\begin{equation}\label{chap1-eq-ecrossb}
{\bf E}=-{\bf V}\times{\bf B}
\end{equation}
for the relation between the electric field ${\bf E}$, fluid velocity ${\bf V}$, and magnetic field ${\bf B}$. These relations must be completed by some equation of state  relation between the plasma pressure $P$, density $N$ and temperature $T$ (in energy units) for which usually the ideal gas law $P=NT$ is taken to hold in its adiabatic version, assuming that shock formation proceeds on such a short scale that the temperature cannot adjust. This is in fact not an unreasonable assumption as the shock cannot be in thermal equilibrium as has been argued above on different reasons.

The three possible shocks in such a case are just the {\it fast}, {\it slow} and {\it intermediate} shocks related \index{shocks!fast}\index{shocks!slow}\index{shocks!intermediate}to the three MHD wave modes that have been mentioned above. They may occur depending on which of the wave phase speeds is exceeded by the flow, and for the slower shocks, of course, under the additional condition that the faster waves are inhibited in the medium because otherwise, when the faster waves would be excited in the interaction of flow and obstacle they would propagate upstream at faster speed than the shock itself and inform the flow about the presence of the obstacle. No shock would be formed in this case from the slower Alfv\'enic or slow modes. Since this is the more realistic case the most frequently observed shocks are fast shocks. However, occasionally also slow or intermediate shocks have been claimed to have been detected in interplanetary space. 

The four possible shock transitions in terms of the relations between flow and MHD wave mode speeds \citep{DeHoffman1950,Balogh2005} are:
\begin{eqnarray}
{\rm trans\, 1} &:&V>c_{ms}^+,  \qquad\qquad\quad
{\rm trans\, 2} :  c_{ms}^+> V> c_{int}  \nonumber\\ [-0.5ex]
\nonumber\\
{\rm trans\, 3}&: &c_{int} > V > c_{ms}^-, ~\,\qquad
{\rm trans\, 4}:  c_{ms}^- > V \nonumber 
\end{eqnarray}\vspace{-0.3cm}

\noindent where the velocities $c_{ms}^\pm$ have been defined in Eq.\,(\ref{chap1-eq-cms}), and $c_{int}=V_A\,\cos\theta$ is the angle-dependent Alfv\'en velocity of the intermediate wave. Not all these transitions can, however, be realized. \cite{Wu1992} have shown that entropy considerations allow only for the transitions $1\to 2, 1\to 3, 1\to 4, 2\to 3, 2\to 4$, and $3\to 4$. Of these the first is a fast mode shock transition, while the transition $3\to 4$ is a slow mode, and the remaining ones are all intermediate shock transitions which together with the slow mode might sometimes exist under the above mentioned restrictions. However, usually an obstacle will excite all three waves together, and then only the fast mode will cause a shock. The observation of slow or intermediate shocks thus requires very special conditions to exist in the plasma. In addition, any shocks will have to respect also the evolutionary condition discussed briefly below.

In the MHD frame one can define simple relations between the parameters of the streaming gas upstream and downstream of the shock, the so-called {\it Rankine-Hugoniot relations}. These relations give a first idea of the conditions at the shock transition. They result from the conservative character of the MHD equations which in fact are conservation laws for the mass flow, momentum flow and energy density. They will be derived in the next chapter. The restriction on them is that the processes which determine the generation of entropy causing the irreversibility of the shock must all be strictly confined solely to the shock transition region. This is not so easy to achieve as it might seem at first glance. The shock transition might be much broader and more extended than the proper shock ramp suggests. Hence the parameter values entering the Rankine-Hugoniot relations are taken correctly only sufficiently far away to both sides of the shock ramp to be sure of not mixing in processes that are dissipative and thus are not contained in the MHD Rankine-Hugoniot conservation laws. The problem is then that the shock surface itself must be extended enough compared with the distance from the shock where the parameter values are taken. In addition its curvature should still be negligible in order not to destroy the assumption of shock planarity involved into the Rankine-Hugoniot relations. 

\subsection{Evolutionarity}\index{shocks!evolutionarity}
\noindent In the above paragraph we mentioned that not all of the six possible shock solutions in MHD can be realized. The actual reason for this lies in the so-called condition of evolutionarity of shock waves, which are based on the hyperbolic nature of the conservation laws which allows wave propagation only if it is in accord with causality \citep{Lax1957}. \index{Lax, P. D.}

For MHD waves with dissipation these conditions have been discussed by \cite[][and others]{Jeffrey1964,Kantrowitz1966,Liberman1986} but also hold in the collisionless regime because causality is a general requirement in nature. Causality in this case means that the drop in speed across a shock (in the wave mode of the shock) must be so large that the normal component of the flow downstream of the shock front falls below the corresponding downstream mode velocity. For a fast shock this implies the following order for the normal flow velocities and magnetosonic velocities to both sides of the fast shock:
\begin{displaymath}
V_{1n}>c_{1ms}^+, \quad {\rm while} \quad V_{2n}>c_{2ms}^+
\end{displaymath}
where the numbers 1, 2 refer to upstream and downstream of the fast shock wave. 

The first condition is necessary for the shock to be formed at all; it is the second condition which (partially) account for the evolutionarity. Otherwise the small fast mode disturbances excited downstream and moving upward towards the shock would move faster than the flow, they would overcome the shock and steepen it without limit. Since this cannot happen for a shock to form, the downstream normal speed must be less than the downstream fast magnetosonic speed. Similar conditions hold for any shock as also for large amplitude shocks. Furthermore, for fast shocks the flow velocity must be greater than the intermediate speed on both sides of the shock, while for slow shocks it must be less than the intermediate speed on both sides. These conditions hold because of the same reason as otherwise the corresponding waves would catch up with the shock front, modify and destroy it and no shock could form. 

\subsection{Coplanarity}\index{shocks!coplanarity}
\noindent Finally, another important observation of MHD shocks relates to the directions of the magnetic field and flows to both sides of the shock front. These directions are not arbitrary. At the contrary it can be shown from the MHD conservation laws respectively from the Rankine-Hugoniot relations that the flow and magnetic field directions in front and behind the shock front in MHD lie in the same plane, i.e. they are coplanar. This property had been realized already by \cite{Marshall1955} and has been discussed in depth by \cite{Kantrowitz1966} and others \citep[e.g.][]{Burgess1995}. For a stationary ideal MHD shock wave with no other wave activity or kinetic processes present outside the shock transition such that dissipation takes place solely inside the narrow shock transition and this transition region can be considered as infinitesimally thin with respect to all other physical scales in the plasma, the electric field in the shock rest frame is strictly perpendicular to the magnetic field, given by  Eq.\,(\ref{chap1-eq-ecrossb}),  the equation
\begin{equation}
\nabla\times{\bf E}=0
\end{equation}
which is the stationary Faraday's law, and the shock normal ${\bf n}$ defined in Eqs.\,(ref{{chap1-eq-shocknormal,chap1-eq-nabla}) yield that the scalar product between ${\bf n}$ and the difference in the tangential components of the magnetic field to both sides vanishes:
\begin{equation}
(V_{n2}-V_{n1})\{{\bf B}_{\rm t2}\times{\bf B}_{\rm t1}\}=0
\end{equation}
The difference in the normal components across the shock does clearly not vanish, such that $V_{n2}\neq V_{n1}$. Hence 
\begin{equation}
{\bf n}\cdot\{{\bf B}_2-{\bf B}_1\}=0
\end{equation}
which implies not only that ${\bf n}$ is normal to the tangential components of the magnetic field on both sides of the infinitesimally thin discontinuity, which would be a trivial conclusion, but also that the two tangential components to both sides are strictly parallel. They may -- and should -- have different lengths but will have same direction across the shock.

Coplanarity does not strictly hold, however, even in MHD. For instance, when the shock is nonstationary, for instance when its width changes  with time (which is not excluded by MHD as all such processes are contained in the internal structure of the dissipation region of width $\Delta$, and this is determined by processes not covered in MHD which assumes that $\Delta­$\,const not being a function of time), the right-hand side in the above Faraday's law does not vanish, and coplanarity becomes violated. Also, if a magnetic wave encounters the shock it will be transformed across the shock. However, its magnetic field components might let the tangential magnetic field rotate at some angle across the shock thereby violating the co-planarity condition. There are also other effects which at a real non-MHD shock violate coplanarity. We will not discuss them at this location. One particular case in MHD is noted in the following subsection.

\subsection{Switch-on and switch-off shocks}
\noindent Parallel shocks in MHD should, theoretically, behave exactly like gasdynamic shocks,  not having any upstream  tangential magnetic field component and should also not have any downstream tangential field. The `tangential field' in this case has `no direction'. 

This conclusion does not hold rigourously, however, since plasmas consist of charged particles which are sensitive to fluctuation in the field and can excite various waves in the plasma via electric currents which then become the sources of magnetic fields. We will later see that kinetic effects in parallel and quasi-parallel shocks play an important role in their physics and are well capable of generating tangential fields at least on scales shorter than the ion scale. 

However, even in MHD one stumbles across the interesting fact that this kind of shocks must have peculiar properties. The reason is that they are not, as in gasdynamics, the result of steepened {\it sound} waves, in which case they would simply be purely electrostatic shocks. At the contrary, the waves propagating {\it parallel} to the magnetic field are {\it Alfv\'en} and {\it magnetosonic} waves. Alfv\'en waves contain transverse magnetic field components. These transverse wave fields are in fact tangential to the shock. Hence, if a purely parallel shock steepens, the transverse Alfv\'en waves do steepen as well, and the shock after the transition from upstream to downstream switches on a tangential magnetic component which originally was not present. Such shocks are called {\it switch-on} shocks. Similarly one can imagine the case that a tangential component behind the shock is by the same process switched off by an oppositely directed switch-on field, yielding a {\it switch-off shock}. Both cases are theoretically possible and models have been provided for instance by \cite{Kennel1988} for the resistive MHD case, even though there is little experimental or observational evidence for the existence of such shocks in space.

The problem of whether or not such shocks exist in MHD is related to the question whether or not an Alfv\'en wave steepens non-linearly when propagating into a shock. To first order this steepening for an ordinary Alfv\'en wave is zero. However, to second order because of its weak transverse magnetic component which is seen by a wave trailing the leading Alfv\'en wave. This trailing wave therefore propagates slightly oblique to the main magnetic field and thus causes a second order density compression which in addition to generating a shock-like plasma compression changes the Alfv\'en velocity locally. In the case when the trailing wave is polarised in the same direction as the leading wave it also increases the transverse magnetic field component downstream of the compression thereby to second order switching on a tangential magnetic component. A whole train of trailing waves of same polarisation will thus cause strong steepening in both the density and tangential magnetic field.

Clearly, this kind of shocks is a more or less exotic case of MHD shocks whose importance is not precisely known \citep[the rare case of observation of a switch-on shock has been reported by][]{Farris1994}. There are other efficient mechanisms of magnetic field generation in shock waves based on the famous Weibel instability which come into play when the shocks have high Mach numbers and the plasma is anisotropic. These mechanisms are particularly strong when relativistic effects must be taken into account which is the case under most astrophysical conditions. These effects are not anymore simple MHD but appear naturally in the kinetic treatment of shocks. In the last chapter of this book we will briefly touch on them.
\begin{figure}[t!]
\hspace{0.0cm}\centerline{\includegraphics[width=0.8\textwidth,clip=]{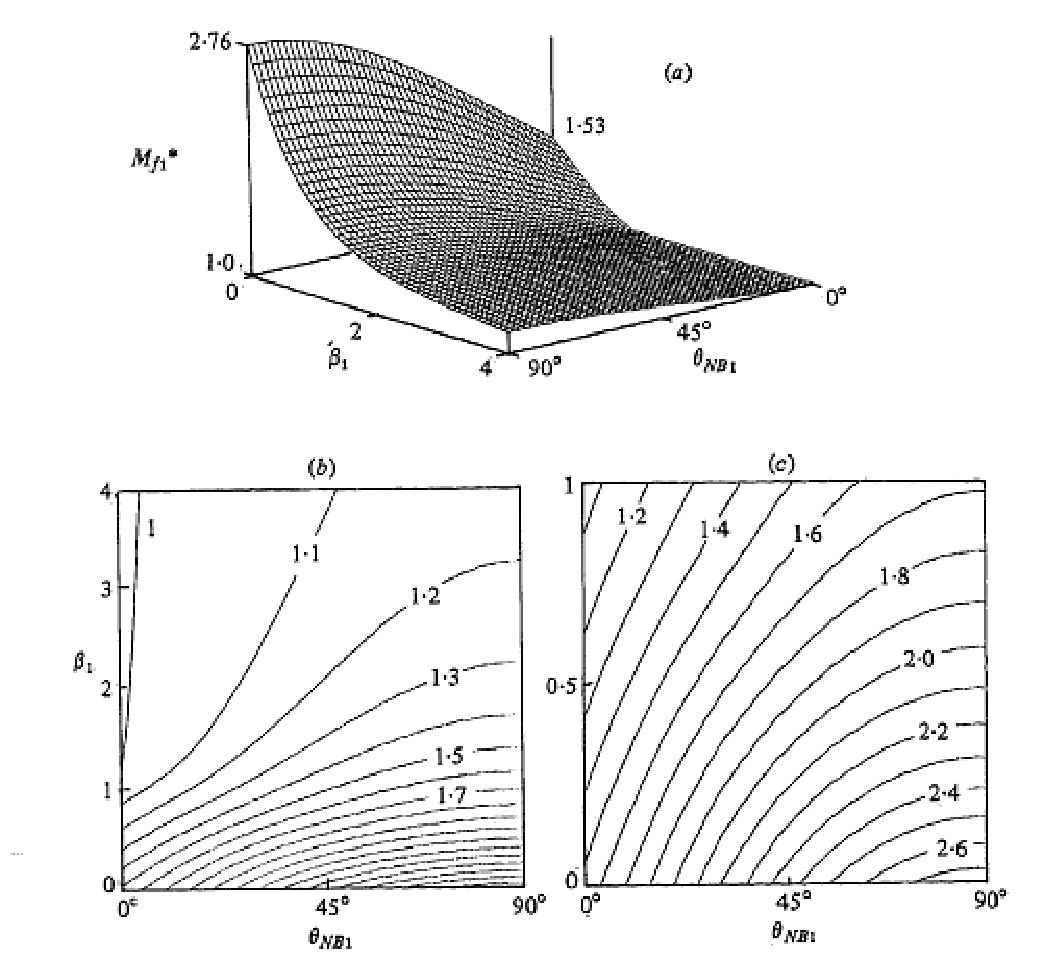} }
\caption[1]
{\footnotesize Parametric dependence of ${\cal M}_c$ for a fast shock on the upstream  plasma-$\beta_1$ (top) and shock angle $\thetabn$ for the special case of adiabatic $\gamma=\frac{5}{3}$. The lower panels show two $\beta_1$  ranges of critical Mach number contours\citep[from][]{Edmiston1986}. For large $\beta_1>1$ the critical Mach number is close to ${\cal M}_c\sim 1$, while for smaller $\beta_1$ it is a strong function of $\thetabn$ having its lowest value ${\cal M}_c= 1.53$ for parallel and ${\cal M}_c=2.76$ for perpendicular shocks. The latter value is the same as that originally inferred already by \cite{Marshall1955}.}\label{chap1-fig-kennel01}
\end{figure}

\section{Criticality}\index{shocks!criticality}
\noindent In this final section of this preparatory chapter we single out a most important property of shocks which leads to another {\it physically} justified classification of collisionless shocks into {\it subcritical} and {\it supercritical} shocks according to their Mach-numbers ${\cal M}<{\cal M}_c$  being smaller or ${\cal M}>{\cal M}_c$ being larger than some critical Mach-number ${\cal M}_c$. 
We already mentioned that at high Mach-numbers the existence of a critical Mach number for collisionless shocks was predicted long ago from consideration of the insufficiency of dissipation in the shock to provide fast enough retardation of the inflow, plasma thermalization, and entropy production. 
For a resistive shock \cite{Marshall1955} had numerically determined the critical Mach number to ${\cal M}_{c}\approx 2.76$. 
 
Subcritical shocks are capable of generating sufficient dissipation to account for retardation, thermalization and entropy in the time the flow crosses the shock front from upstream to downstream. The relevant processes are based on wave-particle interaction between the shocked plasma and the shock-excited turbulent wave fields. These processes will be discussed in Paper 3. 

For supercritical shocks this is, however, not the case. Supercritical shocks must evoke other mechanisms different from simple wave-particle interaction of getting rid of the excess energy in the bulk flow that cannot be dissipated by any classical anomalous dissipation. Above the critical Mach number the simplest way of energy dissipation is rejection of the in-flowing excess energy from the shock by reflecting a substantial part of the incoming plasma back upstream. The physical processes involved into the reflection process and its effects on the structure of the shock will be discussed below in separate chapters for both quasi-perpendicular and quasi-parallel shocks. 

Independent of this the determination of the critical Mach number poses an interesting problem which was posed by \cite{Kantrowitz1966} who realized that the magnetic width of the shock must always exceed the magnetic Reynolds length for dissipative shocks and that the finite magnetic field compression ratio therefore sets an upper limit to the rate of resistive dissipation that is possible in an MHD shock. Plasmas possess several dissipative lengths depending on which dissipative process is considered. Any nonlinear wave propagating in the plasma should steepen so long until its transverse scale approached the longest of these dissipative scales when dissipation sets in and limits its amplitude \citep{Coroniti1970}. 

Thus when the wavelength of the fast magnetosonic wave approaches the resistive length the magnetic field decouples from the wave by resistive dissipation, and the wave speed becomes the sound speed downstream of the shock ramp. The condition for the critical Mach number is then given by $V_{2n}=c_{2s}$. Similarly, for the slow mode shock because of the different dispersive properties the resistive critical Mach number is defined by the condition $V_{1n}=c_{1s}$. Since these quantities depend on wave angle they have to be solved numerically which has been done by \cite{Edmiston1986}. The critical fast mode Mach number varies between 1 and 3, depending on the upstream plasma parameters and flow angle to the magnetic field. It is usually called {\it first} critical Mach number because there is theoretical evidence in simulations for a second critical Mach number which comes into play when the shock structure becomes time dependent \citep{Krasno2002}, whistlers accumulate at the shock front and periodically cause its reformation. The dominant dispersion is then the whistler dispersion. An approximate expression for this {\it second} or {\it whistler} critical Mach number is
\begin{equation}\label{chap-1-eq-critmachnumber}
{\cal M}_{2c}\propto \left(\frac{m_i}{m_e}\right)^\frac{1}{2}\cos\thetabn
\end{equation}
where the constant of proportionality depends on whether one defines the Mach number with respect to the whistler phase or group velocities. For the former it is $\frac{1}{2}$, and for the latter $\sqrt{27/64}$ \citep{Oka2006}. These authors have used Geotail data to confirm the existence of this whistler critical Mach number which separates the regions of sub-critical and super-critical shocks in bulk flow velocity $V_{1n}$/magnetic angle $\thetabn$ space. 

It is clear that it is the smallest critical Mach number that determines the behaviour of the shock. In simple words: ${\cal M}>1$ is responsible for the existence of the shock under the condition that an obstacle exists of the flow is in some way disturbed such that fast waves can grow, steepen and form shocks. When, in addition, the flow exceeds the next lowest Mach number for a given $\thetabn$ the shock at this angle will make the transition into a supercritical shock and under additional conditions, which have not yet be ultimately clarified, will start reflecting particles back upstream. If because of some reason this would not happen the flow might have to exceed the next higher critical Mach number until reflection becomes possible. In such a case the shock would become metastable in the region where the Mach number becomes supercritical, will steepen and shrink in width until other effect and ultimately reflection of particles can set on.
\begin{figure}[t!]
\hspace{0.0cm}\centerline{\includegraphics[height=0.3\textheight,width=1.0\textwidth,clip=]{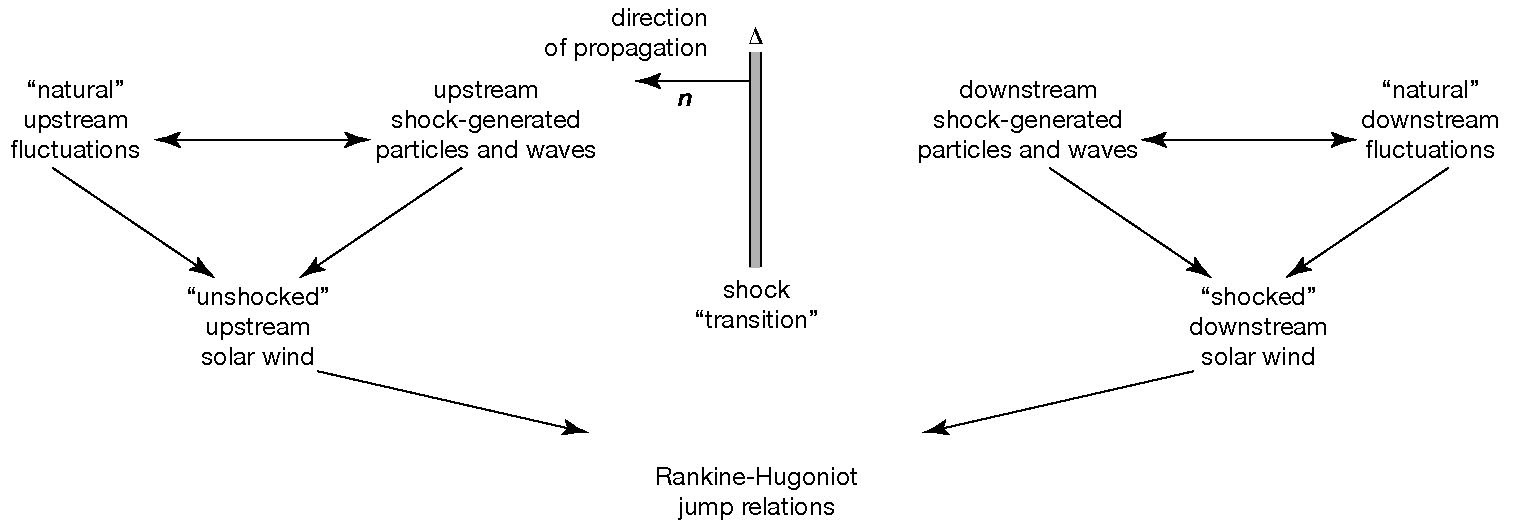} }
\caption[1]
{\footnotesize A schematic representation of the complexity of processes accompanying a collisionless supercritical shock and the relationship between the various states upstream and downstream of the shock transition \citep[after][]{Balogh2005}. There is a noticeable symmetry between the upstream and downstream shock regions, even though the effects and physical properties of both regions are very different belonging to two mutually coupled but different plasma states.}\label{chap1-fig-balogh05}
\end{figure}

\section{Remarks}
\noindent
In this chapter we have not yet dug into the theory and observation of collisionless shocks. Instead we have tried to provide the basic philosophy of shock formation. Recently \cite{Balogh2005} gave a beautiful and comprehensive account of the complexity involved into the shock problem that is illustrated in Figure\,\ref{chap1-fig-balogh05} and which we like to cite at this occasion because we cannot say it in any better way: 
\\[1ex]
\noindent
{ ``...the concept [of shocks] originally inherited from collision-dominated media has been successfully extrapolated to collisionless plasmas ... both qualitatively and quantitatively. There is nevertheless an inherent contradiction at the heart of the concept of collisionless shock waves. On the one hand, a discontinuous transition between two states of the plasma is assumed, represented by nominally well-defined parameters which characterise unambiguously the upstream and downstream states, and which are linked by continuity equations, augmented as necessary by additional equations. On the other hand, the necessary dissipation in the shock transition itself generates phenomena which can propagate information, primarily in the form of particle populations which have interacted with the shock, away from the shock front, upstream as well as downstream in general, and which, in turn, modify in a complex manner the states of the plasma and the temporal and spatial functions of its constituents.... As a result, the collisionless shock possibly never propagates into a region which has not already been modified, to a greater or lesser extent, by phenomena which originate at the shock and which therefore affect, through a set of complex feedback processes, the shock transition itself. This picture is fully born out by numerically simulated shocks, in which the initial conditions, or `states' of the upstream and downstream plasma are fully defined at the outset; but once the shock is generated, the characteristic parameters of the medium evolve as a result of particles propagating away from the shock, generating, through a set of (usually nonlinear) wave particle interactions and a range of instabilities, significantly more complex `states' than was assumed at the start..." }
\\[1ex]
\normalsize
Clearly, the picture developed in this text by \cite{Balogh2005} is that of a supercritical shock, it already refers to the complicated nonlinear set of equations that determine the behaviour of shocks, and it also anticipates much of what will follow in the coming chapters, the particle reflection process from supercritical shocks, the effects those reflected particles have on their surroundings, transportation of information, mass, energy, and fields, excitation of instabilities, waves, and the entire zoo of nonlinear interactions, convection and reaction on the shock which, as we have already clarified, is an entity that is not in thermal equilibrium and therefore evolves and reforms continuously trying to reach thermal equilibrium but being hindered to ultimately establish it by the supercritical inflow. It is no surprise that shocks not only constitute an extraordinarily interesting but also a challenging phenomenon in particular when the plasma is collisionless.

We have in this chapter listed most of the terms and classifications of shocks and the basic terms which are used in their description. We encountered several ways of classification of collisionless shocks which we will follow in the later chapters of this text. 
Their properties, however, must be described by models which depend on the assumption of which scales are considered to be important for the processes to be investigated. 
This will be done in the next Chapter 2 where a number of models are described starting from gasdynamic models and proceeding with increasing sophistication to kinetic and particle models of collisionless shocks.





\end{document}